\documentclass[preprint]{aastex}

\slugcomment{AJ, in press}
\newcommand{\etal}{{\it et al.\/}}

\begin{document}
\title{Observational Limits on a Distant Cold Kuiper Belt}
\author{R. L. Allen\altaffilmark{1}, G. M. Bernstein\altaffilmark{1}}
\affil{Department of Astronomy, University of Michigan,
830 Dennison Bldg., Ann Arbor, MI 48109,
rhiannon, garyb@astro.lsa.umich.edu}

\and
\author{R. Malhotra\altaffilmark{1}}
\affil{Department of Planetary Sciences, University of Arizona,
1629 E.~University Blvd., Tucson, AZ 85721,
renu@lpl.arizona.edu}

\altaffiltext{1}{Visiting Astronomer, National Optical Astronomy
Observatories, which is operated by the Association of Universities
for Research in Astronomy, Inc., under contract to the National
Science Foundation.}

\begin{abstract}

Almost all of the $>600$ known Kuiper belt objects (KBOs) have been
discovered within 50~AU of the Sun.  One possible explanation for the
observed lack of KBOs beyond 50~AU is that the distant Kuiper belt is
dynamically very cold, and thus thin enough on the sky to have slipped
between previous deep survey fields.  We have completed a survey
designed to search for a dynamically cold distant Kuiper belt near the
invariable plane of the Solar system.  In $2.3~{\rm deg}^2$ we have
discovered a total of 33 KBOs and 1 Centaur, but no objects in
circular orbits beyond 50~AU.  We find that we can exclude at 95\% CL
the existence of a distant disk inclined by $i\le1\arcdeg$ to the
invariable plane and containing more than 1.2 times as many $D>185$~km KBOs
between 50 and 60 AU as the observed inner Kuiper belt, if the distant
disk is thinner than $\sigma=1\fdg75$.

\end{abstract}


\section{Introduction}
More than 500 Kuiper belt Objects (KBOs) have been discovered since
the first, 1992 QB1, was detected almost a decade ago. Nearly all of
these are currently within 50~AU of the sun, and only one has been
detected beyond 54~AU, despite surveys sensitive enough to detect
160~km-diameter KBOs to 60~AU. Such surveys have placed strong upper
limits on the density of KBOs in an outer Kuiper belt \citep{ABM01,
TLBE, JLT, Gl01}.  There are several possible explanations for the
observed absence of distant KBOs, each with distinctly different
predictions for the Kuiper belt beyond 50~AU, and distinctly different
implications for the history of the solar system.

In the simplest view of the Kuiper Belt, the region beyond 50~AU,
where the gravitational perturbations of the giant planets become
negligible, has remained dynamically cold, with KBOs moving on their
primordial circular and coplanar orbits. It is inferred that the
region inside 48~AU has been heavily depopulated, by a factor of 10-100 in
mass, over the age of the solar system as a result of planetary
perturbations and mutual collisions \citep{S96, SCb, Le93, Du95}. In
the absence of such perturbations, the outer disk is expected to be a
factor of 5--100 times as dense as the presently observed inner Kuiper
belt.  However, as yet no discovered KBO has been determined to be in
a circular orbit beyond 48~AU.  Shallow surveys, with typical
magnitude limits of $R<23.5$, have searched for KBOs in more than
$100\,{\rm deg}^2$ of sky, but objects at $\ge50$~AU must be
$\ge280$~km in diameter to be detected in such surveys.  (In quoting
sizes of KBOs in this paper, we assume a 4\% albedo.)  It is desirable
to search for smaller objects since the growth timescales for KBOs in
the outer Kuiper belt, though highly uncertain, are likely longer than
in the inner Kuiper belt.  Only a few square degrees, however, have
been surveyed to magnitudes $R\gtrsim24.5$.

\citet{Ha00} has pointed out that if the outer disk were dynamically
very cold, thus restricted to a thin line on the sky, it could have
easily slipped between the previous sparse deep survey fields.  The
most likely location for this distant cold disk would be the
invariable plane, the plane normal to the total angular momentum of
the solar system, which is inclined 1\fdg6 to the ecliptic \citep{Allen76}.
Most deep KBO surveys, including our previously published results
\citep{ABM01}[ABM], have been conducted close to the ecliptic plane,
possibly missing a thin disk near the invariable plane. We have
therefore supplemented the ABM survey with four new fields that are
strategically chosen on the sky to maximize the likelihood of
detecting a thin trans-Neptunian disk near the invariable plane. We
present here new limits on a putative cold disk beyond 50~AU that
result from our observations in these new survey fields combined with
previously published data.

\section{Distant Disk Survey Observations}
The new observations were taken with the KPNO Mayall 4-m telescope and
Mosaic camera, during two consecutive nights in March 2001.  The
observation and reduction methodologies are identical to those of
ABM. In brief: data were taken using a broad-band $VR$ filter similar
to that described in \citet{JLC}. Each field is observed in a
series of 7 to 9 exposures each night, each exposure 480~s long.  The
images were registered and then combined with a ``digital tracking''
technique to search for faint moving objects: after registration, the
images from each night are summed (with sigma-clipping rejection) into
a deep template. This template is then subtracted from each individual
image in the opposing night, after matching the point spread
functions. This removes all stationary objects from each image,
leaving only moving objects and cosmic rays. These blank images are
searched individually for bright KBOs, then combined at potential KBO
motion vectors to search for faint slow-moving objects. The images
were combined tens of thousands of times to search all likely orbits
between 30 and 100~AU.

As the observations were taken in fields very near opposition, the
distance can be estimated to within 10\% from the apparent
velocity. Although we cannot determine orbits with the short
(two-night) observational arc obtained here, a distance alone 
is usually sufficient to exclude the possibility of a distant circular orbit.

We discovered 10 new KBOs in this 2001 survey data, over an effective
area of $1.0~{\rm deg}^2$. The effective area is simply the subtended
survey area multiplied by the detection efficiency---the latter factor
being probability
of detecting a KBO at a particular magnitude. The effective area
is determined by inserting hundreds of artificial moving
objects into the data, then repeating the search process. In our 2001
survey, the detection efficiency drops by 50\% at a magnitude of $R\sim24.9$
due to noise, but has no dependence on velocity or distance as
all KBOs within 200~AU will move appreciably in the two-night
baseline. This limiting magnitude varies from field to field due to
seeing and integration-time variations, while the peak detection efficiency
varies primarily with stellar density and field overlap.  Data for
each field are listed in Table~\ref{fieldtable} and the details for
each discovered object are listed in Table~\ref{objtable}.

Also included in the Tables are details of the fields and detected
objects from the 1998 and 1999 ABM data, which have an effective area
of 1.3 square degrees with mean limiting magnitude of $R=25.4$.
Altogether, our surveys between 1998 and 2001 cover 2.3 square degrees
and we discovered a total of 33 KBOs and 1 Centaur.  The discovered
objects were found at distances between 20 and 53~AU, but none of the
objects beyond 48~AU are on circular orbits.  In our 2001 survey, no
KBOs were found beyond 46~AU.  We have clearly detected no possible
distant cold disk members, even though we could have detected KBOs
larger than $D$=185~km to distances of 55~AU or greater over nearly
the entire survey area with our stated magnitude limits.

\begin{figure}
\plotone{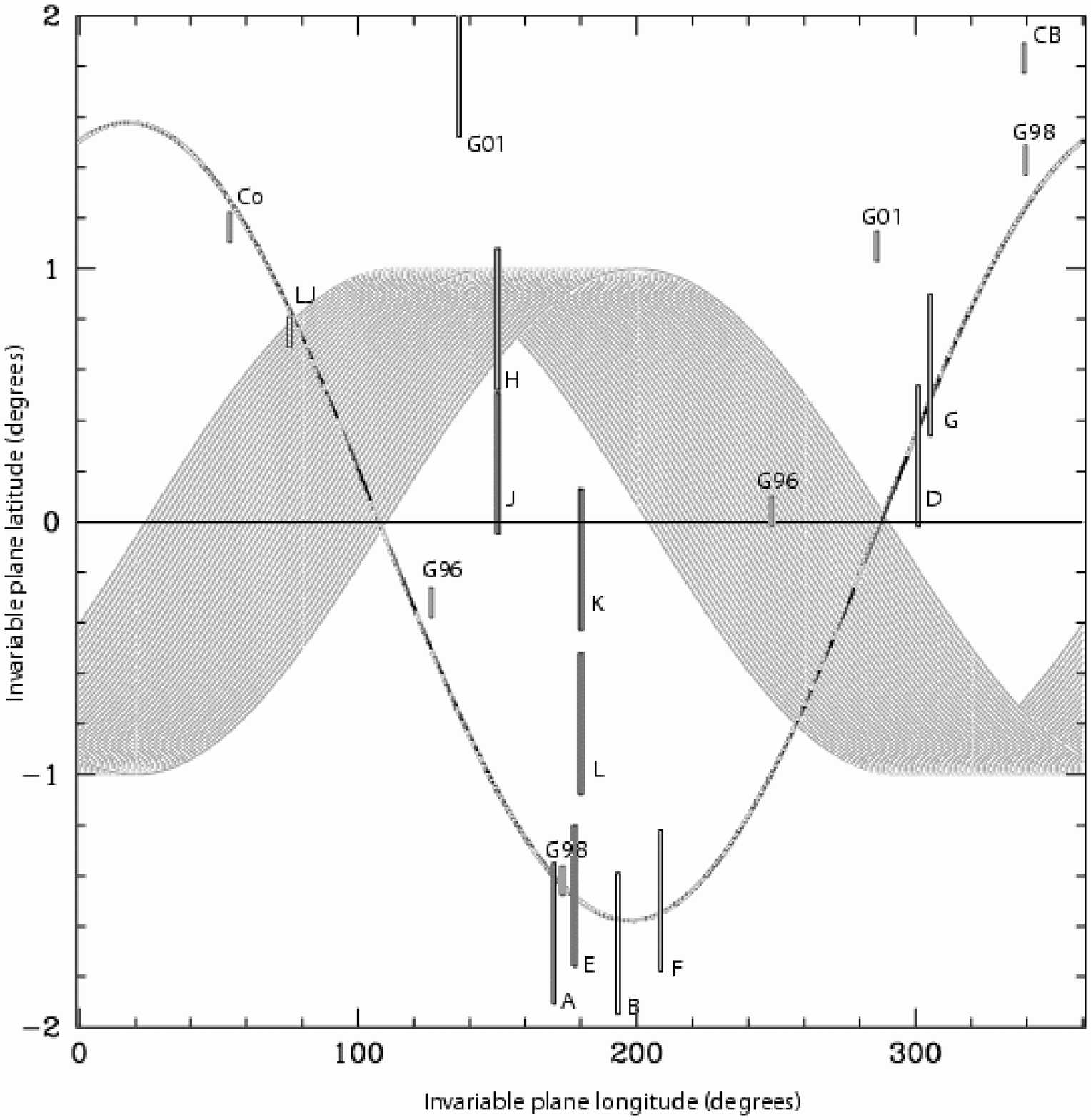}
\caption[invfields.eps]{
Fields from this survey, and others deep enough to provide limits on
the location of distant thin disk at $R=24.8$, are shown here in
invariable plane coordinates. The heavy line shows the location of the
ecliptic plane.  The lighter curves show the locations where a
distant disk, inclined to the invariable plane by 1\arcdeg, might
escape detection in surveys to date.  The distant disk is assumed to
have a Gaussian thickness of 0\fdg5, contain as many $D>185$~km
objects as the ``classical'' Kuiper belt, and be excluded if there is
$<5\%$ chance of being consistent with our observations.
\label{invfields}
}
\end{figure}

Figure~\ref{invfields} shows in invariable-plane coordinates the
extent of the ABM fields (A--G), the new fields (H--L), and other
authors' surveys.  The heavy line is the ecliptic plane (the plane
containing the orbit of the Earth), about which the previous surveys
are clustered.  Observing in the ecliptic rather than the invariable
plane (the total angular momentum plane of the Solar System) has been
the norm, as the difference between the two is not important when
considering a thick disk such as the Kuiper belt inside 50~AU. It is
only when searching for a disk with vertical dispersion
$\sigma\la2\arcdeg$ that this difference becomes significant.

Deep KBO surveys conducted by other researchers \citep{CB99, Gl01,
Gl98} generally cover too little area (on the order of a Keck field
$\sim0.01~{\rm deg}^2$) to provide additional constraints on a thin
disk.  In the analysis described in the next section, we have included
the KBO survey by \citet{Gl01}, conducted with the
Canada-France-Hawaii Telescope.  This survey discovered 1999~DG8, for
which a one-night arc suggests a distance of $\approx60$~AU.  This is
the {\em only} known object which could have a circular orbit beyond
50~AU, but the short arc means that the eccentricity of the orbit is
completely indeterminate---it could easily be a scattered disk member.
For the following analysis we must calculate the probability of cold
disk models being consistent with the observations.  We sum the
probabilities of the model yielding 1 or 0 objects in the \citet{Gl01}
CFHT field, since the nature of 1999~DG8 is indeterminate.  This,
along with the high latitude of the field [1.5 to 2 deg to the
invariable plane], substantially weaken its ability to constrain a
cold disk in the invariable plane.

\section{Limits on a Distant Disk}

We did not detect any KBOs beyond 50~AU in the primordial, circular
orbits expected of objects in a distant cold disk.  We wish to
constrain the total number of KBOs which could be resident in such a disk.
We will consider only KBOs with diameter $D\ge185$~km, since such
objects are detectable at $>50\%$ efficiency in all our fields at a
distance of 55~AU, assuming 4\% albedo.  
We express the total population $N_{\rm cold}$ of $D>185$~km objects
between 50 and 60~AU as $f N_{\rm CKBO}$, where $N_{\rm CKBO}$ is the
number of classical KBOs (CKBOs) larger than
$D=185$~km. \citet{TLBE} determine $N_{\rm CKBO}\sim3500$.  The cold
disk is assumed to have some inclination $i$ and ascending node
$\Omega$ relative to the invariable plane.  The surface density of
cold disk objects on the sky is assumed to have a Gaussian
distribution in the vertical direction with dispersion $\sigma$, and
to be uniform with disk longitude.  An exponential vertical
distribution yields similar results for most disk widths.

We ask then what is the probability $P(O|f,i,\Omega,\sigma)$ of a
given disk model yielding the observations $O$.  For simplicity we
assume that all the $N_{\rm cold}$ objects in the 50--60~AU range have
the apparent magnitude $m=24.8$ that a $D=185$~km object would have at
a distance of 55~AU.  This is conservative in the sense that many of
the objects would presumably be larger and brighter, but we do not
want to assume any particular size distribution for the cold-disk
members.  Note that for $f=1$, we may imagine taking the CKBO
population, collapsing (cooling) it down to a thin disk, then moving
the disk out from its present 38--48~AU annulus to a 50--60~AU
annulus.  If the ``classical'' belt is truly a depleted and excited 
inner portion of the full Kuiper Belt, then this is a lower bound to the
expected distant disk.

For chosen disk parameters $\{f,i,\Omega,\sigma\}$, we convolve the
model sky surface density with the survey geometry and detection
efficiencies specified in Table~\ref{fieldtable}, yielding an expected
number of detections.  The expected number of detections for each
field, $N_i$, for each cold disk orientation is then
\begin{equation}
N_i = 3550 \,\frac{E_i}{2}\,\frac{\Delta{\rm (RA)} \cos{\rm (dec)} }{2\,\pi}\,
\left |\,{\rm erf}\,(\frac{|b_{i1}|}{\sqrt{2}\,\sigma}) - {\rm erf}\,(\frac{|b_{i2}|}{\sqrt{2}\,\sigma})\,\right |,
\end{equation}
where $E_i$ is the detection efficiency in each field, $\Delta{\rm
(RA)}$ is the field width and $b_{i1}$ and $b_{i2}$ are the top and
bottom field latitudes in the cold disk plane.  Since the actual
number of detections is zero (save possibly for 1999~DG8 in the CFHT
field), $P(O)$ is easily calculated from Poisson statistics.  A given
disk configuration is excluded at 95\% confidence if $\ge3$ detections
are expected.

As an example, 74\% of the nodes of a very thin disk ($\sigma=0\fdg5$)
inclined at $i=1\fdg0$ to the invariable plane, are ruled out at the
$>95$\% confidence level for $f=1$. Most of the remaining
un-excluded disks travel through our field H, due to its slightly brighter
limiting magnitude. The $i=1\arcdeg$, $\sigma=0\fdg5$ disks which are 
{\em not} ruled out by
our deep fields are plotted in Figure~\ref{invfields}.

\begin{figure}
\plotone{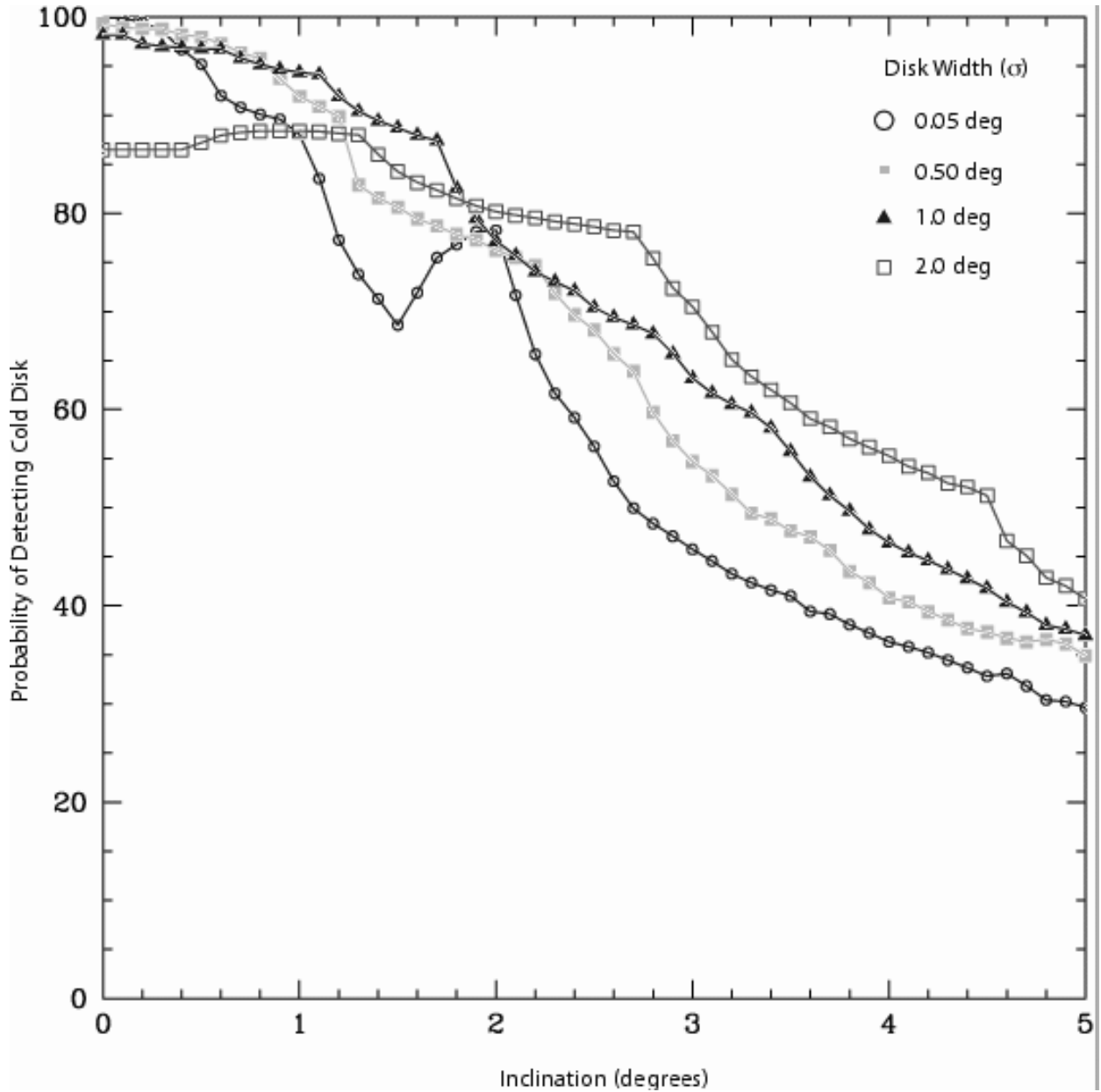}
\caption[disksout.eps]{The probability of detecting a distant disk
with varying thickness and inclination (averaged over ascending nodes)
to the invariable plane. This assumes that the number of
objects larger than $D=185$~km is similar to the number of $D>185$~km
KBOs in the classical Kuiper belt.
\label{disksout}
}
\end{figure}

To simplify the interpretation, we marginalize over the ascending node
$\Omega$ by assuming a uniform prior distribution in the interval
$0<\Omega<2\pi$:
\begin{equation}
P(O|f,i,\sigma) = {1 \over 2\pi} \int_0^{2\pi}P(O|f,i,\Omega,\sigma)\,d\Omega.
\end{equation}
Figure~\ref{disksout} plots the probability $P(O|f=1,i,\sigma)$. As we
allow the cold disk model to be more inclined to the invariable plane,
our limits become weaker because these disks more easily dodge the
surveyed fields.  Our limits also become weaker as the model disks
become thicker, at inclinations $\le1\arcdeg$, since the sky density
of the KBOs drops in our fields.  This situation reverses at higher
inclinations, where the thicker disks ($\sigma>0.5\arcdeg$) are more
likely to be detected, as they are unable to slip between our fields
and the inclination variation matters less.

\begin{figure}
\plotone{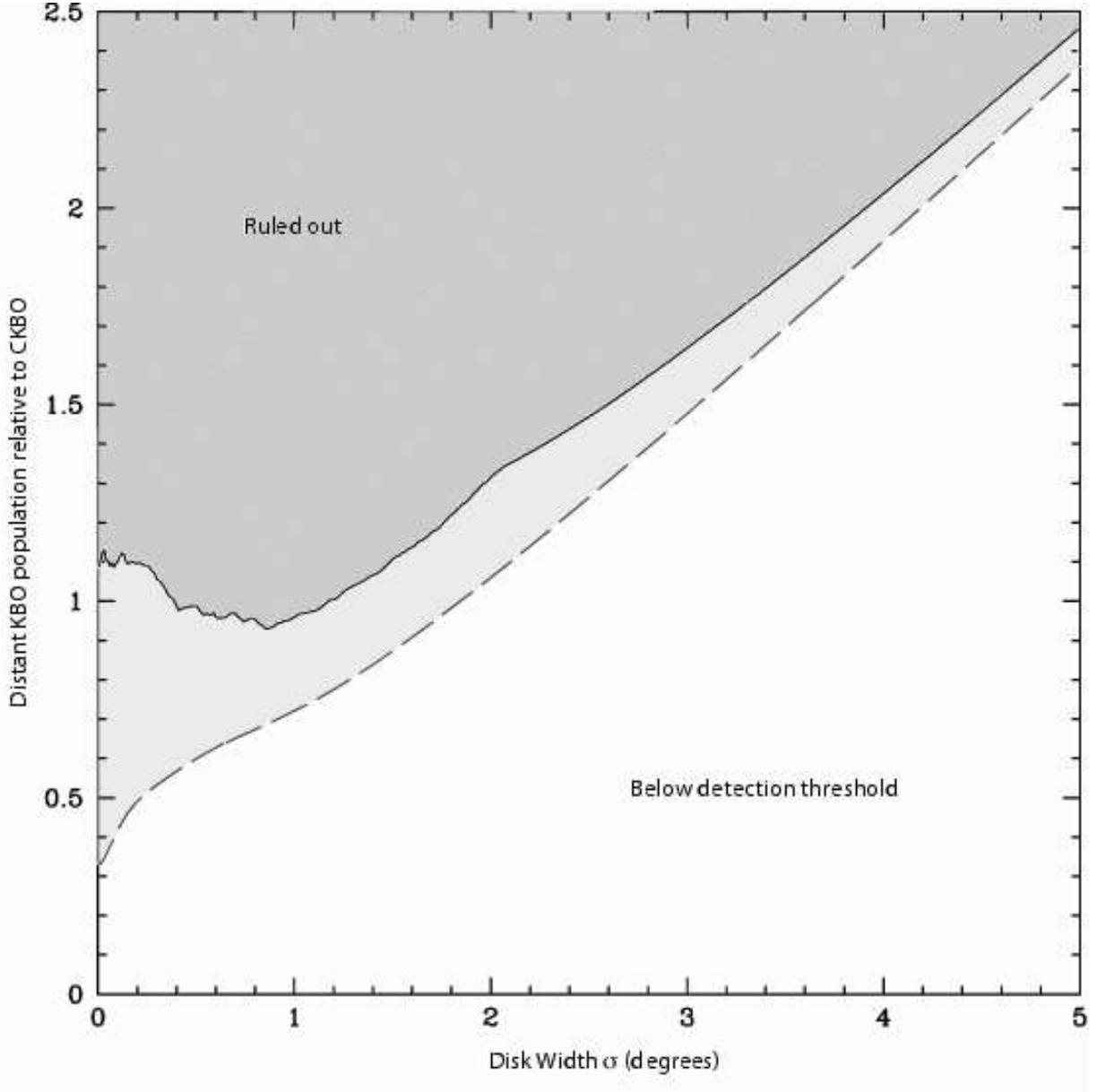}
\caption[diskwidths.eps]{Assuming the distant cold disk is in the
invariable plane ($i=0\arcdeg$), we calculate the upper limit to the
ratio $f$ of $D>185$~km objects relative to the classical Kuiper Belt,
shown by the dashed line.  By marginalizing over the inclination
(averaged over inclination pole angular area) we can also calculate
the upper limit to the ratio $f$ for all cold disks with
$i\le1\arcdeg$ to the invariable plane, shown by the solid line. Disk
populations above each line are inconsistent at $\ge95\%$ CL with our
non-detection.
\label{diskwidths}
}
\end{figure}

If we assume that the distant cold disk must be located in the
invariable plane, then we can place further limits on how many objects
larger than 185~km could exist between 50--60~AU without being
detected in our survey.  Figure~\ref{diskwidths} plots the value of
$f$ at which $P(O|f,i=0\arcdeg,\sigma)=0.05$, as a function of disk
thickness $\sigma$.  A thin disk ($\sigma=0.22\arcdeg$) must have less
than half as many $D>185$~km KBOs as the classical Kuiper belt
($f<0.5$ at 95\% CL). An $i=0\arcdeg$ distant disk with the same number
of objects as the inner belt must have $\sigma>1.85\arcdeg$ to have a
$>5\%$ chance of escaping detection in the surveys to date.  The
apparent width of the classical Kuiper belt is $\approx5\arcdeg$
\citep{TLBE}. If the outer disk has the same width instead of being
dynamically colder, an increase of factor $f>2.36$ over the classical
belt population is ruled out at the 95\% CL.  The case of equivalent
inner- and outer-disk sky distributions in considered in more detail
in ABM.

By also marginalizing over the inclination, we can place similar limits
on $f$ for cold disks within $1\arcdeg$ of the invariable plane. Assuming
that the model disk pole locations are uniformly distributed in 
polar area, then 
\begin{equation}
P(O|f,i\le1\arcdeg, \sigma) = {\int_0^1\, P(O|f,\sigma,i)\,i\, di \over \int_0^1 i\, di}. 
\end{equation}
We can then find $f$ at which $P(O|f, i\le1,\sigma)=0.05$, plotted in
Figure~\ref{diskwidths} as the solid line. We can place tighter limits
on a disk in the invariable plane, as above, than on any disk in any
inclination up to $1\arcdeg$ because more of our fields were located
in the invariable plane. However, we find for $\sigma<1.75\arcdeg$ 
we can exclude at 95\% CL or better the existence of a distant 
disk at any inclination $i\le1\arcdeg$ to the invariable plane that
contains at least 1.2 times as many $D>185$~km KBOs as the Classical 
Kuiper belt. 

\section{Conclusions}
We have provided strong limits on the existence of a distant cold
disk. The simple expectation of a dense thin disk in the invariable plane
appears to be ruled out. However, it is possible that a thin disk composed
of objects smaller than $D=185$~km, undetectable in much of our survey,
is present. This would require an extreme change in the size
distribution of KBOs just beyond $\sim$50~AU, as objects larger than
$D\approx1000$~km have been found at distances near 50~AU. It is not clear
what would cause such a large and sudden change in the size distribution.

Other explanations for the observed lack of distant KBOs involve
dropping the density of objects beyond 50~AU.  One possibility is that
the outer Kuiper belt was dynamically excited by a stellar encounter
early in the history of the solar system \citep{Id00}.  An increase in
the mean eccentricity and inclination of distant KBOs, thus lowering
their apparent sky density and possibly halting accretion of large
KBOs, would make a distant disk much harder to detect. Dynamical
excitation by a stellar encounter results in a distinctive orbital
distribution of KBOs, and could yield limits on the birth cluster
environment of the solar system \citep{AL01}.

It is also possible to explain the lack of KBOs beyond 50~AU if the
primordial planetesimal disk ended at this distance. If Neptune and
Uranus have migrated significantly over the age of the solar system
\citep{Ma93, Ma95, Th99}, the primordial solar nebula surface density
beyond $\sim30$~AU could have been very small initially. This could
suggest that the larger KBOs actually formed interior to 30--40 AU and
were displaced to their present orbits by a large scale rearrangement
of orbits in the outer Solar system.
Circumstellar disks around other stars have been observed to
have diameters ranging between 50--1000~AU \citep{McOd96, BrownD00}. 
A truncation of the solar nebula near 50~AU would fit easily within
this range, with important implications for planetary formation models
of our Solar System.

In order to distinguish between these theories and further constrain the
distant KB population, deeper and wider survey observations of the Kuiper belt
(to limiting magnitudes of $R>25$) will be necessary.
Based on the limits we have found here,
a survey covering approximately 20 square degrees near
the invariable plane should be able to detect a distant disk
dynamically excited to a thickness of 10\arcdeg\ with even half the
number of objects as the CKB. This would constrain the density and
distribution of the Kuiper belt at distances beyond 50~AU, including
the Scattered Disk Objects, enabling a better comparison between our
solar system and extrasolar systems.

\acknowledgements
We thank L. Strolger and M. Holman for attempting to recover some of these
objects, and the staff at Kitt Peak National Observatory for their excellent
support. This work is supported by NASA Planetary Astronomy grant
\#NAG5-7860; GB is further supported by grant \#AST-9624592 from the
National Science Foundation; RM is further supported by NASA 
grants \#NAG5-10346 and \#NAG5-11661.

\clearpage

\thispagestyle{empty}

\begin{deluxetable}{clrrcccccccc}
\rotate
\tablecolumns{11}
\tablewidth{0pt}
\tablecaption{Field Information}
\tablehead{
\colhead{Field} & \colhead{Dates of} & \colhead{$RA$} & \colhead{Dec} &
\multicolumn{2}{c}{Ecliptic Coords.} &
\colhead{Invariable} & \colhead{Exp Time} &
\colhead{Filter} & \colhead{$E_o$\tablenotemark{b}} & \colhead{$m_{50}$\tablenotemark{b}} & \colhead{w} \\
 & \colhead{Observation\tablenotemark{a}} 
& \multicolumn{2}{c}{(J2000)} & \colhead{Long.} & \colhead{Lat.} &
\colhead{Lat.} & \colhead{(s)} & & \colhead{(\%)} & 
}
\startdata
A & 5/19/98             & $11^h\ 37^m\ 34^s$ &  +2 13 20  &
174.0   &  -0.19  &  -1.63  & 12$\times$480 & $R$ & \nodata  & \nodata & \nodata \\
  & 5/28/98 - 5/29/98 * & $11^h\ 37^m\ 22^s$ &  +2 14 34  &
173.9   &  -0.18  &  -1.63  & 19$\times$480 & $R$ & 73  & 24.9 & 0.19\\

B & 5/19/98             & $13^h\ 02^m\ 36^s$ &  -6 46 00  &
197.0   &  -0.03  &  -1.61  & 18$\times$480 & $R$ & \nodata  & \nodata & \nodata \\
  & 5/28/98 - 5/29/98 * & $13^h\ 02^m\ 00^s$ &  -6 42 26  &
196.9   &  -0.09  &  -1.67  & 24$\times$480 & $R$ & 75  & 25.2 & 0.15\\

D & 5/19/98             & $20^h\ 27^m\ 52^s$ &  -19 20 00  &
304.6   &  -0.22  &   0.24  & 18$\times$480 & $R$ & \nodata  & \nodata & \nodata  \\
  & 5/28/98 - 5/29/98 * & $20^h\ 27^m\ 35^s$ &  -19 19 47  &
309.1   &  -1.40  &  0.81  & 26$\times$480 & $R$ & 65  & 25.4 & 0.22 \\

E\tablenotemark{c} & 5/10/99 - 5/11/99 * & $12^h\ 05^m\ 00^s$ &  -0 30 00  & 
181.4   &  0.04   &  -1.48  & 33$\times$480 & $VR$ & 78  & 25.4  & 0.37 \\
  & 5/18/99 - 5/19/99 * & $12^h\ 04^m\ 34^s$ &  -0 32 28  &
181.3   &  -0.01  &  -1.53  & 33$\times$480 & $VR$ & 78  & 25.4  & 0.37 \\

F & 5/10/99 - 5/11/99 * & $14^h\ 00^m\ 00^s$ &  -12 12 00  &
212.2   &   0.03  &  -1.50  & 30$\times$480 & $VR$ & 88  & 25.9 & 0.23 \\
  & 5/18/99 - 5/19/99 * & $13^h\ 59^m\ 21^s$ &  -12 15 15  &
212.1   &  -0.08  &  -1.61  & 33$\times$480 & $VR$ & 88  & 25.9 & 0.23 \\

G & 5/10/99 - 5/11/99   & $20^h\ 45^m\ 00^s$ &  -18 00 00  & 
308.8    &   0.05  &   0.62  & 48$\times$480 & $VR$ & \nodata   &
\nodata & \nodata \\
  & 5/18/99 - 5/19/99 * & $20^h\ 45^m\ 00^s$ &  -18 00 00  &
308.8    &   0.05  &   0.62  & 44$\times$480 & $VR$ & 74  & 25.7 & 0.28 \\

H & 3/27/01 - 3/28/01 * & $10^h\ 24^m\ 35^s$ & 12 01 15 &
153.5  	&    1.93  &  0.80 & 14$\times$480 & $VR$ & 95 & 24.7 & 0.33 \\

J & 3/27/01 - 3/28/01 * & $10^h\ 24^m\ 02^s$ & 11 27 37 &
153.6  &   1.36  &  0.23  & 17$\times$480 & $VR$ & 93 & 24.8 & 0.34 \\

K & 3/27/01 - 3/28/01 * & $12^h\ 15^m\ 23^s$ & -00 09 53 & 
183.6 & 1.38  & -0.15  & 15$\times$480 & $VR$ & 95 & 24.9 & 0.27 \\

L & 3/27/01 - 3/28/01 * & $12^h\ 14^m\ 10^s$ & -00 44 11 &
183.5 &  0.73  & -0.80  & 18$\times$480 & $VR$ & 85 & 24.9 & 0.24 \\ 

\enddata
\tablenotetext{a}{Observations marked with asterisks were searched for
KBOs; unmarked observations were used only for (p)recovery.}
\tablenotetext{b}{$m_R$ and $E_o$ are fit to the detection efficiency $E$
for detection of implanted KBOs using $E={{E_o} \over 2}  {\rm
erfc}[(m-m_{50})/\sqrt{2}w].$
$E_o$ is the peak detection efficiency for each field, 
$m_{50}$ is the point where the efficiency
is 50\% of $E_o$, and $w$ is a width parameter. }
\tablenotetext{c}{Detection efficiencies for fields E and F are based
on averages from nights 1 \& 2 and nights 3 \& 4, as these fields 
search the same section of sky.}
\label{fieldtable}
\end{deluxetable}
\thispagestyle{empty}

\clearpage

\begin{deluxetable}{ccccccrccc}
\rotate
\tablewidth{0pt}
\tablecaption{Objects Discovered}
\tablehead{
\colhead{MPC} & \colhead{Field} & \colhead{Arc} &
\colhead{$m_R$\tablenotemark{a}} &
\colhead{$a$\tablenotemark{b}} & \colhead{$e$\tablenotemark{b}} & \colhead{$i$} &
\colhead{Heliocentric} & \colhead{Diameter\tablenotemark{c}}  &\colhead{H\tablenotemark{d}}\\
\colhead{Designation} & & \colhead{Length} & &
\colhead{(AU)} & &
\colhead{(\arcdeg)} &
\colhead{Dist. (AU)} &
\colhead{(km)} &
}
\startdata
1998 KD66 & B & 10 d & 24.7 & 
\nodata & \nodata & $ 6.4\pm 2.9 $ & $42.9\pm3.7$ & 117 & 8.4 \\
1998 KE66 & B & 10 d & 25.0 & 
\nodata & \nodata & $ 2.5\pm 0.9 $ & $41.0\pm3.4$ & 94 & 8.9 \\
1998 KF66 & B & 10 d & 24.5 & 
\nodata & \nodata & $ 6.7\pm 1.5 $ & $31.8\pm2.0$ & 70 & 9.5 \\
1998 KG66 & B & 10 d & 25.1 & 
\nodata & \nodata & $ 3.5\pm 1.5 $ & $45.2\pm4.1$ & 109 & 8.5 \\
1998 KY61 & D & 42 d & 23.7 & 
$44.1\pm 0.1$ & $0.05\pm0.10 $ & $ 2.1\pm 0.0 $ & $46.5\pm0.0$ & 220 & 7.0 \\
1998 KG62 & D & 2 opp & 22.9 &
$43.4\pm 0.0$ & $0.05\pm0.01 $ & $ 0.8\pm 0.0 $ & $45.3\pm0.0$ & 301 & 6.3 \\
1998 KR65 & D & 2 opp & 22.9 &
$43.5\pm 0.0$ & $0.02\pm0.00 $ & $ 1.2\pm 0.0 $ & $44.4\pm0.0$ & 289 & 6.4 \\
1998 KS65 & D & 2 opp & 23.7 &
$43.7\pm 0.0$ & $0.03\pm0.00 $ & $ 1.2\pm 0.0 $ & $42.3\pm0.0$ & 181 & 7.4 \\
1999 JV127 & E & 8 d & 23.7 & 
$18.2\pm 0.2$ & $0.15\pm0.08 $ & $19.2\pm 0.7 $ & $20.9\pm0.3$ & 43 & 10.5 \\
1999 JA132 & E & 9 d & 23.9 & 
$42.0\pm 3.8$ & $0.07\pm0.13 $ & $ 7.3\pm 0.7 $ & $45.2\pm1.1$ & 189 & 7.3 \\
E2-01\tablenotemark{e} & E & 1 d & 24.9 &
\nodata & \nodata & $7.1\pm3.5$ & $31.7\pm4.2$ & 58 & 9.9 \\
1999 JB132 & F & 8 d & 23.5 & 
\nodata & \nodata & $17.1\pm11. $ & $39.1\pm3.7$ & 170 & 7.6 \\
1999 JC132 & F & 1 d & 24.3 & 
\nodata & \nodata & $ 5.4\pm 2.1 $ & $39.0\pm2.6$ & 117 & 8.4\\
1999 JD132 & F & 2 opp & 23.6 & 
$45.4\pm3.3$ & $0.22\pm0.16$ & $10.5\pm 0.0 $ & $42.8\pm0.2$ & 198 & 7.3 \\
1999 JE132 & F & 9 d & 24.1 & 
$32.4\pm 5.0$ & $0.20\pm0.22 $ & $29.8\pm 6.4 $ & $39.1\pm1.4$ & 129 & 8.2 \\
1999 JF132 & F & 9 d & 24.0 & 
\nodata & \nodata & $ 1.6\pm 0.4 $ & $43.1\pm2.1$ & 164 & 7.7 \\
1999 JH132 & F & 9 d & 25.5 & 
\nodata & \nodata & $ 0.6\pm 0.2 $ & $41.1\pm2.4$ & 75 & 9.4 \\
1999 JJ132 & F & 9 d & 25.5 & 
\nodata & \nodata & $ 3.2\pm 1.2 $ & $50.1\pm2.8$ & 111 & 8.5 \\
1999 JK132 & F & 9 d & 24.8 & 
\nodata & \nodata & $16.0\pm 5.9 $ & $39.0\pm2.5$ & 93 & 8.9 \\
1999 KT16  & F & 1 d & 25.1 & 
\nodata & \nodata & $ 8.5\pm 3.7 $ & $46.3\pm3.1$ & 114 & 8.4 \\
1999 KK17  & F & 9 d & 25.0 & 
\nodata & \nodata & $ 8.7\pm16. $ & $49.8\pm7.5$ & 139 & 8.0 \\
1999 KL17  & G & 90 d & 25.2 &
$46.2\pm 0.2$ & $0.03\pm0.17 $ & $ 2.8\pm 0.0 $ & $47.6\pm0.0$ & 115 & 8.4 \\
1999 KR18  & G & 89 d & 24.9 &
$43.3\pm 0.3$ & $0.21\pm0.03 $ & $ 0.6\pm 0.0 $ & $52.6\pm0.1$ & 162 & 7.7 \\
G3-01\tablenotemark{e}  & G & 9 d & 25.6 &
$39.9\pm2.7$  & $0.16\pm0.05$ & $1.6\pm0.1$ & $33.4\pm1.4$ & 47 & 10.4 \\
2001 FB185 & H & 1 d & 24.6 & \nodata & \nodata & $4.5\pm1.5$ & $39.8\pm3.1$ & 106 & 8.7 \\
2001 FC185 & J & 1 d & 23.8 & \nodata & \nodata & $21.2\pm8.9$ & $39.0\pm3.4$ & 147 & 7.9 \\
2001 FE193 & K & 1 d & 23.2 & \nodata & \nodata & $3.1\pm1.1$ & $42.8\pm2.5$ & 234 & 6.9 \\
2001 FG193 & K & 1 d & 24.2 & \nodata & \nodata & $3.4\pm1.6$ & $41.8\pm2.5$ & 141 & 8.0 \\
2001 FD193 & K & 1 d & 24.2 & \nodata & \nodata & $12.3\pm4.6$ & $44.7\pm2.7$ & 161 & 7.7 \\
2001 FC193 & K & 1 d & 24.8 & \nodata & \nodata & $2.8\pm1.5$ & $46.4\pm2.6$ & 132 & 8.2 \\
2001 FJ193 & L & 1 d & 24.5 & \nodata & \nodata & $1.1\pm0.6$ & $36.4\pm2.4$ & 93 & 8.9 \\
2001 FF193 & L & 1 d & 23.8 & \nodata & \nodata & $2.0\pm1.6$ & $44.7\pm2.6$ & 194 & 7.3 \\
2001 FH193 & L & 1 d & 24.8 & \nodata & \nodata & $8.9\pm3.5$ & $40.9\pm2.6$ & 102 & 8.7 \\
2001 FL193 & L & 1 d & 24.9 & \nodata & \nodata & $1.0\pm0.6$ & $40.9\pm2.5$ & 97 & 8.8 \\
\enddata
\tablenotetext{a}{All 1999 and 2001 objects were detected in the $VR$ filter.
The listed magnitude may be transformed to standard $R$ via $R=VR -
0.46(V-R-0.5)$.  As $0.3\lesssim V-R \lesssim0.7$ for KBOS
\citep{Te00}, the $R$ mag varies by 0.2~mag or less with color.}
\tablenotetext{b}{No data are given when the arc is too
short to provide meaningful constraint.}
\tablenotetext{c}{Diameters assume albedo of 0.04.}
\tablenotetext{d}{Absolute magnitude in $R$}
\tablenotetext{e}{Objects E2-01 and G3-01 were not reported to the MPC
due to insufficient S/N on the recovery observations.}
\label{objtable}
\end{deluxetable}

\end{document}